\documentclass[conference]{IEEEtran}
\usepackage{xcolor}
\usepackage{amsmath, amsfonts, amssymb, epsfig, adjustbox}
\usepackage{graphicx}
\usepackage{mathtools}
\usepackage{hyperref}
\usepackage{cite}
\usepackage{algorithm}
\usepackage{algorithmicx}
\usepackage[noend]{algpseudocode}
\usepackage{comment}
\usepackage[switch]{lineno}
\usepackage{enumitem}
\usepackage{multirow}
\usepackage{mathtools}
\usepackage[amsthm,  thmmarks,amsmath]{ntheorem}

\begin{document}
\onecolumn
\copyright Personal use of this material is permitted. Permission from IEEE must be obtained for all other uses, in any current or future media, including reprinting/republishing this material for advertising or promotional purposes, creating new collective works, for resale or redistribution to servers or lists, or reuse of any copyrighted component of this work in other works.

This paper has been accepted by The 32nd International Conference on Computer Communications and Networks (ICCCN 2023)

\twocolumn
\newpage
\title{Data-Driven Next-Generation Wireless Networking: Embracing AI for Performance and Security}
\author{
    \IEEEauthorblockN{Jiahao~Xue\IEEEauthorrefmark{1}, Zhe Qu\IEEEauthorrefmark{3}, Shangqing~Zhao\IEEEauthorrefmark{2}, Yao~Liu\IEEEauthorrefmark{1} and Zhuo~Lu\IEEEauthorrefmark{1}}
    \IEEEauthorblockA{
      \IEEEauthorrefmark{1}University of South Florida, Tampa, FL, USA.\\
	  \IEEEauthorrefmark{2}University of Oklahoma, Norman, OK, USA.\\
	  \IEEEauthorrefmark{3}Central South University, Changsha, Hunan, China. \\      
    }
}

\maketitle

\begin{abstract}
	New network architectures, such as the Internet of Things (IoT), 5G, and next-generation (NextG) cellular systems, put forward emerging challenges to the design of future wireless networks toward ultra-high data rate, massive data processing, smart designs, low-cost deployment, reliability and security in dynamic environments. As one of the most promising techniques today, artificial intelligence (AI) is advocated to enable a data-driven paradigm for wireless network design. In this paper, we are motivated to review existing AI techniques and their applications for the full wireless network protocol stack toward improving network performance and security. Our goal is to summarize the current motivation, challenges, and methodology of using AI to enhance wireless networking from the physical to the application layer, and shed light on creating new AI-enabled algorithms, mechanisms, protocols, and system designs for future data-driven wireless networking. 	
\end{abstract}
\begin{IEEEkeywords}
Wireless network, AI, Machine learning, Performance, Security.
\end{IEEEkeywords}
\vspace{-0.3cm}

\section{Introduction}

The rapid advancement of wireless technology leads to a revolution in daily life. The deployment of Internet of Things (IoT) \cite{ye2019deep,santos2016intelligent}, intelligent networking \cite{merchant2018deep,wang2019data}, cloud computing \cite{xiong2021warmonger,xu2017deep}, 5G and next-generation (NextG) cellular networks \cite{mehrabi2019decision,sadeghi2017optimal} make new demands on the capabilities for efficient and secure network operations. Conventional methods for wireless networking have been generally based on theoretical models, pre-defined operational procedures, or empirical guidelines \cite{hampton2013introduction,yang2015fifty}. Considering the complicated structures and operational protocols of modern wireless networks, conventional methods may not be always efficient, robust, or secure in handling dynamic network operational environments with massive data exchange \cite{hampton2013introduction,shlezinger2020deepsic,guo2020convolutional}. 

Recently, the wide applications of artificial intelligence (AI) and machine learning have drawn increasing attention in the area of wireless networking. New research areas have already emerged to apply AI techniques to enhance wireless network performance and security \cite{teng2019low,xu2020application,xu2021deep,wang2018efficient,han2017two,kuadey2021deepsecure,chinchali2018cellular,shah2020multiagent}. In particular, network operations generate various data of large volume. Without relying on specific mathematical modeling or operational guidelines, AI techniques have enabled a data-driven paradigm to process wireless signals and network traffic in an efficient and secure manner. For example, AI has been adopted in different network layers to improve the network throughput, communication efficiency, and reduce energy consumption and various costs \cite{teng2019low,xu2020application,xu2021deep}; and many system designs have also embraced AI to enhance the confidentiality, integrity, and availability of wireless networks \cite{wang2018efficient,han2017two,kuadey2021deepsecure}. 

In this paper, we aim to provide an overview of existing applications of data-driven AI in the wireless network. We study them from two perspectives: performance and security, and discuss the advantages of using data-driven AI approaches compared with conventional approaches toward wireless network performance and security. In particular, we discuss the following major topics in this paper. 

\begin{itemize}
	\item We classify existing popular AI techniques into supervised learning, unsupervised learning, and reinforcement learning, and briefly introduce common algorithms associated with them. 
	\item We comprehensively present the use of AI techniques to improve the performance and security in wireless network designs throughout the full protocol stack. We begin with the physical (PHY) and medium access control (MAC) layers, which are the main focuses of the recent AI-enabled research. Then, we summarize substantial efforts that have recently applied AI techniques to mechanisms at the network layer and above. 
	\item Based on the state-of-the-art, we discuss what the challenges lie on the path ahead in adopting and creating AI techniques for future wireless network designs. 
\end{itemize}

The remaining sections of this paper are organized as follows. In Section~\ref{sec:sum}, we briefly summarize AI techniques. In Sections \ref{sec:low} and \ref{sec:high}, we discuss the use of AI for wireless networking in lower layers (PHY and MAC) and higher layers (network layer and above), respectively. We summarize the future challenges of AI for wireless networking in Section~\ref{sec:challenges} and conclude this paper in Section~\ref{sec:end}.

\section{Brief Summary of AI techniques} \label{sec:sum}
\begin{figure} 
	\centering
	\includegraphics[width=3.2in]{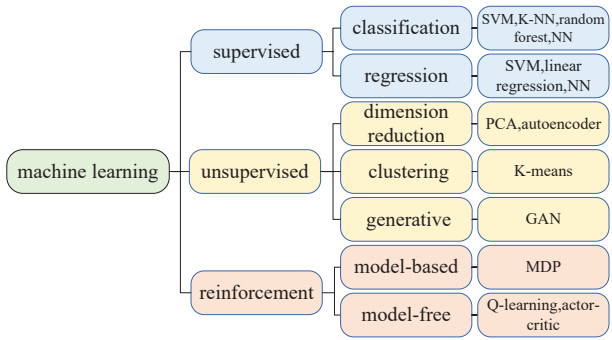}
	\caption{Taxonomy of AI/machine learning techniques, along with commonly used models.}
	\vspace{-0.3cm}
	\label{fig:ml}
\end{figure}

Before we discuss AI techniques for wireless network designs, we briefly introduce and classify AI and machine learning techniques. Fig.~\ref{fig:ml} shows the classification of machine learning techniques into three main categories: supervised learning, unsupervised learning, and reinforcement learning \cite{mahesh2020machine,zhang2020taxonomy}, along with common algorithms in each category.

\begin{itemize}
	\item Supervised learning involves techniques that are trained by explicit labels. Supervised learning includes classification and regression algorithms. Common algorithms include support vector machines (SVM), K-nearest neighbors (K-NN), random forest, linear regression, neural network (NN) based deep learning such as feedforward neural network (FNN), recurrent neural network (RNN), and convolutional neural network (CNN) \cite{aggarwal2018neural}. 
	
	\item Unsupervised learning does not need labeled data, which is classified into dimension reduction, clustering, and generative algorithms. Principal component analysis (PCA) and autoencoder are two common dimension reduction algorithms. Autoencoder has a similar nature to wireless communication because it has an encoding-decoding structure \cite{aoudia2019model}. K-means is a widely used clustering algorithm. Unlike discriminative classification, generative adversarial network (GAN) is a generative machine learning algorithm \cite{aggarwal2018neural}.
	
	\item Reinforcement learning is categorized into model-based and model-free algorithms. One of the most common models for model-based reinforcement learning is based on the Markov decision process (MDP). Model-free algorithms are categorized into value-based algorithms such as the Q-learning, and policy-based algorithms such as the actor-critic algorithm. Besides, deep reinforcement learning is an algorithm that combines reinforcement learning with deep learning. Multi-agent reinforcement learning enables multiple agents in the environment. 
\end{itemize}

\section{AI in PHY and MAC Layers} \label{sec:low}

In a wireless network, the lower layers, including the PHY layer and the MAC layer, are responsible for interacting with the spatially and temporally varying wireless medium to ensure efficient, reliable, and secure wireless communication. Studies have demonstrated that machine learning designs have been successfully integrated into lower-layer designs, enabling wireless networks to (i) adapt to fluctuating environmental conditions (e.g., signal propagation, attenuation, interference) \cite{dorner2017deep,raj2020design, goutay2020deep,he2020model,tan2018low,chaudhari2020reliable, balevi2020massive, lu2020multi, yang2020machine}, and (ii) enhance security against various threats, such as the identification of unauthorized access to wireless networks, suspicious behavior, or protecting the confidentiality, integrity, and availability of wireless networks \cite{senigagliesi2020comparison,fang2018learning,xiao2016phy,xiao2017phy,senigagliesi2019statistical, jiang2018virtual,benzaid2016intelligent }. In this section, we elaborate on how existing approaches integrate machine learning into wireless networks from two aspects: (i) improving the performance and (ii) enhancing the security. 

\subsection{Using AI for Performance}
\begin{figure}
	\centering
	\includegraphics[width=3.35in]{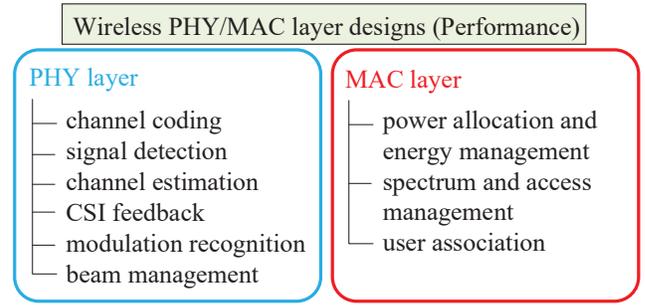}
	\caption{Machine learning for different mechanisms to improve the PHY and MAC performance}
	\vspace{-0.3cm}
	\label{fig:lowlayer1}
\end{figure}

Fig.~\ref{fig:lowlayer1} summarizes how AI has been applied to different designs at the PHY and MAC layers to improve the communication performance. 

\subsubsection{Improving PHY Layer Performance}
In the PHY layer, we discuss how machine learning can be utilized to 1) optimize channel coding, 2) detect high-dimensional signals, 3) advanced channel estimation, 4) optimize CSI feedback procedures, 5) detect modulation without decoding, and 6) improve beam management.

\vspace{0.05cm}\noindent{\bf{Channel Coding:}}
Channel coding is an essential technique to improve the reliability of wireless communication over a noisy channel (e.g., mitigating wireless collisions \cite{zhao2020comb}). Some studies have used machine learning to design advanced coding processes \cite{teng2019low,nachmani2016learning,dorner2017deep,raj2020design,aoudia2019model}. For example, the work of \cite{teng2019low} introduced an RNN-based decoder for polar codes in 5G radio. RNN decomposes the iterative operations of the conventional decoder into layers that can significantly reduce memory consumption by sharing the weights of different iterations. Deep learning can train a Tanner graph for error-correcting codes \cite{nachmani2016learning}. The deep learning framework improves the performance of the belief propagation decoding algorithm with little extra complexity. Autoencoders have also gained broad attention \cite{dorner2017deep,raj2020design,aoudia2019model}. By using autoencoder in the deep generative model, \cite{raj2020design} was able to reconstruct the Gray coding before decoding by using the prior information obtained from the channel model.

In addition, machine learning-based modulation and coding schemes for link adaptation were proposed in \cite{huang2021deep,saxena2021reinforcement}. Parameters for modulation and coding scheme have a probabilistic model based on signal-to-interference-plus-noise ratio (SINR) \cite{saxena2021reinforcement}. This model is formulated as a multi-armed bandit problem under the reinforcement learning framework. Efficient solutions are used to learn the optimal parameters given the channel state.

\vspace{0.05cm}\noindent{\bf{Signal Detection:}}
The conventional signal detection method based on the maximum-likelihood estimation can be an NP-hard problem \cite{yang2015fifty}. Recent advances in multiple-input multiple-output (MIMO) technology with high-dimensional signals have even exacerbated the complexity problem at the receiver. To address this issue, existing research has focused on developing AI-based methods for detecting MIMO signals \cite{goutay2020deep,he2020model,tan2018low,chaudhari2020reliable,he2018model,khani2020adaptive,shlezinger2020deepsic}. The authors in \cite{khani2020adaptive} proposed a deep learning-based MIMO detection called MMNet. MMNet aims to learn the parameter models of an iterative decoder, which eliminates the need to make an impractical assumption that one knows the MIMO channel matrix distribution. In MMNet, parameters can be adaptively adjusted by measuring the channel matrix continuously. Many existing studies on signal detection always assumed that the channel is linear with perfect channel state information (CSI). However, in practice, this assumption does not always hold. The work in \cite{shlezinger2020deepsic} replaced the traditional iterative detection algorithm with deep learning to enhance resilience against CSI error and channel non-linearity.

\vspace{0.05cm}\noindent{\bf{Channel Estimation:}}
In wireless networks, channel interference generally incurs a negative impact, particularly in MIMO systems. The interference can significantly reduce the accuracy of channel estimation. Deep learning has been used for improving the channel estimation performance in \cite{he2018deep,liu2020deep,mehrabi2019decision,balevi2020massive}.
While minimum mean square error estimation is the most accurate, it has a high level of complexity, whereas least squares estimation is faster but less accurate. To combine the advantages of both methods, a deep learning method has been proposed in \cite{balevi2020massive}, which theoretically proved that noise can be effectively filtered so that least squares estimation can approach the close performance of minimum mean square error estimation. The work in \cite{mehrabi2019decision} considered a 5G vehicular network where conventional methods use Doppler rate estimation to estimate decision-directed channels, but these methods do not work well in a highly dynamic environment. The work proposed to use deep learning to learn a channel without knowing the exact Doppler rate, enabling more accurate decision-directed channel estimation.

\vspace{0.05cm}\noindent{\bf{CSI Feedback:}}
It is necessary to perform the sounding process in a beamforming-based multiuser MIMO system, in which each wireless station feedbacks its CSI to the access point for precoding to mitigate interference across different stations. It has been shown that the use of compressed sensing in deep learning can further improve the efficiency of CSI feedback \cite{wang2018deep,wen2018deep,guo2020convolutional,lu2020multi}.
The CNN-based CsiNet+ framework in \cite{guo2020convolutional} has an encoder-decoder structure to compress and quantize the CSI matrix. As opposed to the traditional quantization in deep learning, which requires retraining when changing the bit quantization rate, CsiNet+ is trained by optimizing quantization offset, thus CNN parameters can be fixed without retraining. In \cite{lu2020multi}, a deep learning framework is proposed for extracting CSI features at different resolutions. CSI matrices with different densities require different kernels and resolutions. To extract features at different resolutions, two different convolutional layers are applied in parallel.

\vspace{0.05cm}\noindent{\bf{Modulation Recognition:}}
Automatic modulation recognition (AMR) is a term used to describe the identification of the modulation scheme used in a communication system without decoding signals. Recently, AI techniques have shown their promising potential in this application \cite{wang2019data,lin2020contour}. The work in \cite{aslam2012automatic} used K-NN combined with genetic programming to identify four common modulation schemes. K-NN evaluates the fitness of new features generated by genetic programming based on input features. Due to the simplicity of K-NN, the design is low in complexity without compromising the classification accuracy. It is also possible to recognize signal waveforms by transforming complex-valued signals into contour stellar images, then using deep learning methods for image recognition \cite{lin2020contour}, in which the amplitude, frequency, phase, noise, and error are represented by colors and shapes. Simulation results in \cite{lin2020contour} also validated that such computer vision technology can be applied to AMR.

\vspace{0.05cm}\noindent{\bf{Beam Management:}}
In 5G/NextG wireless networks, millimeter-wave (mmWave) has been used to support higher data rate transmissions. Due to the directional nature of the mmWave technology, each device uses a dedicated beam to communicate with its connected peer. However, this leads to a complex beam management procedure between the transceivers. Machine learning methods have been proposed for solving a variety of beam management problems. For example, considering the beam selection in a vehicle-to-vehicle network, the dynamic nature of such a network makes it difficult to find a beamforming solution that can accommodate its changes \cite{ahn2022machine,yang2020machine,klautau20185g}. The work in \cite{yang2020machine} uses iterative SVM to classify beamforming and select the optimal one. Iterative SVM uses signal power, path loss, and angle of arrival/departures (AoA/AoD) as features for model training, and predicts the analog beam when the link between vehicles is changed.


Tracking beams in a dynamic network is also challenging \cite{chiang2021machine,lim2021deep,sarkar2021machine}. For beam tracking in an unmanned aerial vehicle (UAV) system, drones need to quickly switch beam directions to maximize the SINR when they fly around. Therefore, a fast beam tracking technique is necessary. Although it is difficult to obtain an accurate channel model in a UAV system, Q-learning can learn from the tracking experience without a model to predict tracking \cite{chiang2021machine} by optimizing beam selection by using the SINRs from different beams as rewards.


Beam alignment aims to find and maintain the optimal beam direction between transceivers \cite{heng2021machine,ma2020machine,domae2021machine}. Due to the small antennas used in 5G devices, conventional beam alignment techniques may not be feasible for small devices. The work in \cite{heng2021machine} uses two machine learning classifiers, i.e., random forest and multilayer perceptron (MLP), for beam alignment. Given a user's location, the work uses exhaustive search to find optimal access points and beamforming, then uses locations as features to train the classifiers. After training, classifiers only need the user's location to predict optimal access points and beamforming. The classifiers are shown robust to the general urban outdoor environment. 

\subsubsection{Improving MAC Layer Performance}
Machine learning has been used at the MAC layer to optimize the performance by managing a variety of resources as follows. 

\vspace{0.05cm}\noindent{\bf{Power Allocation and Energy Management:}}
Several studies have adopted AI methods for power allocation and energy management \cite{meng2018deep,mismar2019framework,lee2019resource,zhang2020power,nasir2019multi,xiao2019reinforcement,liu2018deepnap,xu2017deep}. The work in \cite{zhang2020power} considered a cognitive radio network that consists of sensors, primary users, and secondary users. The primary and secondary users share the same spectrum resource. Primary users can adjust their power allocation based on their rules. However, secondary users can not obtain primary users' power allocation information and have to use the strength of the received signal from sensors to change their power allocation. As a result, \cite{zhang2020power} designed a deep reinforcement learning framework for secondary users to predict primary users' transmission power allocation. In \cite{liu2018deepnap}, a deep Q-network was developed to learn the optimal sleeping rules for mobile networks to reduce energy consumption. In the proposed deep Q-network method, data traffic from different time periods can be effectively learned, thus reducing the bias caused by current traffic. The method has been shown stable and adaptable in a dynamic environment than conventional Q-learning.

\vspace{0.05cm}\noindent{\bf{Spectrum and Access Management:}}
Spectrum and channel access management can also leverage AI to improve its efficiency \cite{wang2018deepmultichannel,naparstek2018deep,challita2018proactive,li2019multi,koushik2017intelligent,xu2020application,li2020deep, ye2020deepnoma,zhong2021ai,wang2021joint, rupasinghe2015reinforcement,nisioti2019robust,nisioti2019fast}. For example, non-orthogonal multiple access (NOMA) has become a popular design for 5G/NextG networks, which requires comprehensive management of power and spectrum, such that the receiver can successfully decompose signals from users. A multi-task deep learning-based NOMA was proposed in \cite{ye2019deep}, which is able to modulate, spread symbols, and detect. The design is to create a new structure of autoencoder. Each user's bits are modulated to a symbol independently by one of the isolated sublayers. The symbol is spread to a sequence, and then multiple sequences are jointly detected by a neural network. The design was further improved in \cite{ye2020deepnoma} by introducing a balancing technique among users to avoid some users getting trapped in local optima. Targeting the dynamic spectrum access scenario where wireless devices dynamically and autonomously access and use available spectrum resources in a given frequency band, \cite{naparstek2018deep} considered a probability model in multichannel wireless networks. In the model, each user accesses a channel to send data packets with a probability, and will be informed whether the packets are received successfully. A multi-agent deep reinforcement learning was created to learn the best time slot for spectrum access and maximize the data rate on the channel. The study investigated several cases with different rewards and objective functions, including cooperative rewards and global rewards.

\vspace{0.05cm}\noindent{\bf{User Association:}}
User association is a process to associate a user with an appropriate access point based on geographical, channel, interference, and bandwidth information. Machine learning techniques for user association have been investigated in \cite{sana2019multi,zhao2018deep,zhao2019deep}. In particular, \cite{sana2019multi} proposed a multi-agent reinforcement learning model to optimize the association decision. In this model, each user is associated with an agent, and the SINR is used to evaluate the goodness. The experimental results show that this model can achieve up to $99.8\%$ of the optimal performance. In \cite{zhao2019deep}, the user association and resource allocation were considered jointly in a large-scale heterogeneous cellular network. When users are associated with different base stations, both network resources and communication quality can be optimized. It is assumed that users do not know the network environment and are selfish to benefit themselves, which is formulated as a stochastic game. By defining the network utility as the reward of each user, their multi-agent deep reinforcement learning framework can find the Nash equilibrium among users. 

\subsection{Using AI for Security}
\begin{figure}
	\centering
	\includegraphics[width=3.35in]{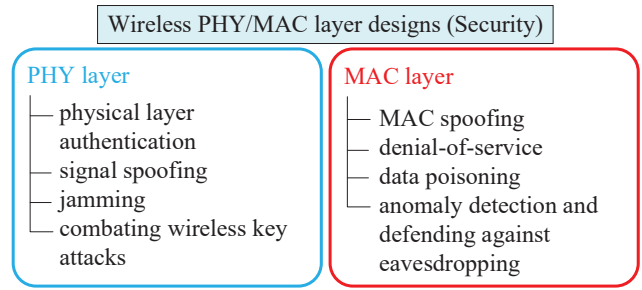}
	\caption{PHY and MAC security mechanisms leveraging AI.}
	\vspace{-0.3cm}
	\label{fig:lowlayer2}
\end{figure}

In addition to using AI to improve the communication performance, the literature also investigated how to leverage AI to enhance wireless network security against various attacks. Fig.~\ref{fig:lowlayer2} summarizes related major topics at the PHY and MAC layers, which will be discussed in the following.

\subsubsection{Security at PHY Layer}
At the PHY layer, various machine learning-based mechanisms have been proposed to enhance authentication, combat spoofing attacks and jamming, and detect anomaly and eavesdropping. 


\vspace{0.05cm}\noindent{\bf{Physical Layer Authentication:}}
It has been demonstrated that secret information can be coupled with random channel responses for secure information exchange without using conventional cryptography. Typically, these designs mainly involve physical-layer authentication (PLA). Some implementations of machine learning on PLA were discussed in \cite{senigagliesi2020comparison,fang2018learning,xiao2016phy,xiao2017phy,senigagliesi2019statistical,merchant2018deep}. For example, \cite{xiao2017phy} used logistic regression to predict unique features in the channel matrix as a way for user authentication, showing better performance than using the conventional received signal strength indicators. In \cite{senigagliesi2019statistical}, a method was proposed to allow Bob using Alice's packets to train one class nearest neighbor algorithm. Packets not classified as belonging to Alice are marked as suspicious. The work in \cite{merchant2018deep} developed a CNN-based radio frequency fingerprinting model by using baseband error signals in the time domain. This method utilizes the frequency offset as a feature during the training process, which is difficult to spoof and therefore can be used to identify attackers.


\vspace{0.05cm}\noindent{\bf{Signal Spoofing Attacks:}}
Spoofing attack is a common attack to compromise the authentication process \cite{hu2021smartphone, shi2020generative, xiao2015spoofing, wang2018efficient}. There are studies specifically targeting spoofing attacks in wireless networks based on AI techniques. For example, the authentication scheme in \cite{wang2018efficient} models the virtual channels of a MIMO system. The sparsity and total energy of users' virtual channels are considered features used by a logistic regression classifier to distinguish spoofing attacks. As the spoofing attacker can be smart and try to learn from waveform and channel status information to improve the spoofing success probability, \cite{shi2020generative} proposed a GAN model that allows the spoofing attacker trains deep learning to obtain the best signals against the defense mechanism obtained by training another deep learning model. This GAN-based attacker can generate signals that are easily misidentified as normal users.

\vspace{0.05cm}\noindent{\bf{Jamming:}} 
Jamming is a common strategy of sending wireless signals with the same frequency in order to disrupt ongoing communication \cite{alagil2016randomized}. Some machine learning-based anti-jamming methods have been developed in \cite{wu2011anti,labib2015colonel,xiao2015user,erpek2018deep,gwon2013competing,aref2017multi,han2017two}. In \cite{han2017two}, the attacker's goal is to disrupt secondary users in a cognitive radio network. Secondary users leverage spatial diversity to transmit signals at different locations to avoid attackers. This study proposes to use deep reinforcement learning to learn the optimal location for the secondary user at each time slot. The work in \cite{wu2011anti} investigated a multi-channel cognitive radio network where secondary users' access is not protected, thus making them vulnerable to jamming attacks. The secondary user's defense strategy is to switch channels in order to hide from the attacker when the attacker is searching for different channels. In this study, the channel hopping is modeled as a Markov decision process (MDP) where the transition probability describes the action of the secondary user.

\vspace{0.05cm}\noindent{\bf{Combating Wireless Key Attacks:}}
The varying wireless channel state can be leveraged to generate a random secret key. A defense method against wireless key attack was considered in \cite{letafati2021deep}. In a wireless network, the wireless secret key generation technique enables key agreement protocols to ensure safe encryption. The performance of wireless secret key generation can be evaluated by the secret key rate. However, both hardware impairment and the forged signal can downgrade the secret key rate. Secret key generation requires randomness distillation that uses pilot signals, thus attackers can inject forged pilot signals. Hardware impairment leads to the mismatch of randomness observation, which can be fixed by deep learning. The attacker is defended against by using RNN to predict the source of common randomness and enhance the randomness distillation. The defense method in \cite{letafati2021deep} has up to $30\%$ improvement compared with others.

\subsubsection{Security at MAC Layer}
At the MAC layer, we discuss how AI techniques have been used in security topics related to spoofing, data poisoning, denial-of-service (DoS), and eavesdropping.

\vspace{0.05cm}\noindent{\bf{MAC Spoofing:}}
Spoofing attacks at the MAC layer have been studied for years \cite{fang2016mimicry}, such as using machine learning \cite{jiang2018virtual,benzaid2016intelligent}. When two packets are sent from different MAC addresses, the proposed deep learning classifier in \cite{jiang2018virtual} can identify whether MAC addresses are associated with the same device by CSI even when two devices of the same model are sending messages at the same location and their CSI still has variances. The work in \cite{benzaid2016intelligent} uses sequence numbers of frames associated with identifies features to train a machine learning model. The experiment conducted in a real-world environment shows it is effective in noisy IEEE 802.11 networks.


\vspace{0.05cm}\noindent{\bf{DoS Attacks:}}
DoS attacks targeting the MAC layer are discussed in \cite{agarwal2015detection} to undermine the frame formatting and flow control. The study showed that attackers can flood forged IEEE 802.11 management frames in WiFi. Management frames are essential for the initialization of WiFi setup. A forged management frame can de-authenticate and disconnect devices. Without upgrading protocol or hardware, machine learning-based classifiers can classify de-authentication frames based on the traffic features, such as the number of different frames and their exchange.

\vspace{0.05cm}\noindent{\bf{Data Poisoning:}}
Data poisoning attacks have been proposed in \cite{luo2020adversarial,luo2020attackers,luo2021low} to circumvent multi-access mechanisms. In particular, in a cooperative spectrum sensing scenario, in which sensing devices can send their results to a data fusion server to determine whether a channel is free. Malicious devices can send poisonous data to the fusion center, which may lead to the server making incorrect decisions. Different from traditional statistics-based methods, this line of research has developed surrogate models based on adversarial machine learning for attackers to mimic the fusion center's decision process, based on which to generate poisonous data in a precise way. Experimental results show that the success probability of adversarial machine learning-based attacker achieves up to $82\%$ attack success rate. 

\vspace{0.05cm}\noindent{\bf{Anomaly Detection and Defending against Eavesdropping:}}
Anomaly detection is a method against malicious access or anomalous phenomena \cite{kavousi2020machine, han2021anomaly}. In \cite{kavousi2020machine}, an anomaly detection algorithm for a wireless sensor network implemented in a microgrid is considered. The algorithm adopts machine learning to detect data integrity with a low false alarm rate during the experiments. The study in \cite{satam2020wids} trained a machine learning model with the traffic features under IEEE 802.11 protocol to detect an anomaly. Detecting eavesdropping \cite{meng2019revealing, fang2018no} is a challenge because it is a passive attack and does not need to actively transmit signals. Some anti-eavesdropper defense strategies were developed in \cite{he2018learning,vashist2019securing}. The idea in \cite{he2018learning} is to mix signals with artificial noise to confuse any eavesdropper. FNN is used to optimize the secrecy throughput, which is evaluated by the power of artificial noise power, the time taken by transferring power, and the redundancy of wiretap code. 

\section{AI in Network Layer and Above}\label{sec:high}
\begin{figure}
	\centering
	\includegraphics[width=3.2in]{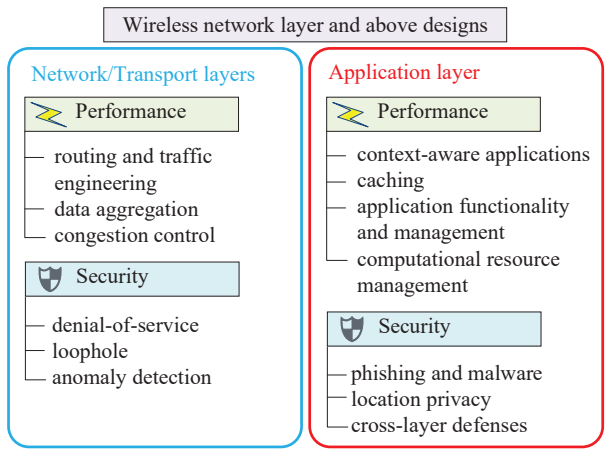}
	\caption{Machine learning to improve performance and security of mechanisms at the network layer and above.}
	\label{fig:midlayer}
	\vspace{-0.3cm}
\end{figure}

While the PHY and MAC layers are always the focus of wireless network research, considerable efforts have also been devoted to using AI for the wireless network layer and above. In this section, we aim to summarize such research efforts toward improving the wireless performance and security. Fig.~\ref{fig:midlayer} summarizes existing machine learning-based mechanism designs toward improving the performance and enhancing security at the network layer and above.

\subsection{AI for Performance}
We first review existing methods of applying AI techniques to improve the network performance. 

\subsubsection{Network and Transport Layers}
Existing studies have been focused on AI-enabled routing, traffic engineering, and data aggregation at the network layer and intelligent congestion control at the transport layer. 

\vspace{0.05cm}\noindent{\bf{Routing and Traffic Engineering:}}
Routing is one of the major tasks in the network layer. Leveraging machine learning methods can help routers determine when and where the data traffic should be sent efficiently \cite{chinchali2018cellular, testi2018machine}. For example, the work in \cite{chinchali2018cellular} considered that an IoT network serves High Volume Flexible Time (HVFT) applications. HVFT needs to transfer a large volume of data to the cloud server such as prefetching videos with ultra-high bit rate. A deep reinforcement learning-based policy was created to coordinate HVFT with other time-sensitive applications such as video streaming for the IoT network. HVFT is scheduled to avoid time-sensitive applications by deep reinforcement learning with the reward set to be the total HVFT throughput. This design is able to transmit $14.7\%$ more data without downgrading time-sensitive applications. 

\vspace{0.05cm}\noindent{\bf{Data Aggregation:}}
Data aggregation can improve the efficiency of wireless networks by reducing redundant data \cite{ye2009optimal,yu2011adaptive}. Conventional aggregation techniques may not be flexibly efficient as they were generally built on fixed routes. In \cite{yu2011adaptive}, a reinforcement learning-based data aggregation design was created for a mobile vehicular network (VANET) scenario. Every vehicle in the VANET uses distributed MDP model to learn from nearby vehicles' actions and rewards. Every vehicle adds a delay before transmitting data as action. The reward is the distance between data in different route nodes. Therefore, data from different vehicles can arrive at the same time and then be aggregated, achieving a good trade-off between delay and redundancy with the number of redundant data reduced without causing a long delay.

\vspace{0.05cm}\noindent{\bf{Congestion Control:}}
A variety of machine learning algorithms have been applied for congestion control at the transport layer, including K-means \cite{taherkhani2016centralized}, SVM \cite{gholipour2017hop}, neural network \cite{al2011end}, and reinforcement learning \cite{li2018qtcp,donta2020congestion,cui2020improving,he2021deepcc, xie2019adaptive,xu2021deep}. Reinforcement learning has received more attention recently. For example, \cite{li2018qtcp} considered a mobile network with varying link bandwidths. Users can switch between links with different channel capacities, which leads to a large-scale dynamic reinforcement learning state space. The congestion window size is defined as the action and network throughput as the reward. Then, Kanerva coding in reinforcement learning is applied to speed up the convergence rate by adequately choosing a part of the state space to approximate the full space. 

\subsubsection{Application Layer}
We briefly discuss common applications where machine learning techniques have been proposed to improve the performance. 

\vspace{0.05cm}\noindent{\bf{Context-aware Applications:}}
Context-aware applications can adapt to serve users based on the context of users. Various machine learning-based applications have been proposed in \cite{yao2017deepsense,santos2016intelligent,li2019smartloc,wang2016device,vashist2020indoor}. For example, \cite{yao2017deepsense} proposes a location-based mobile computing application by using deep learning. It can predict users' tracking or user identification based on biometric motion. Global interactions are obtained by merging local interactions from different sensing modalities. Features such as frequency are extracted to train deep learning. It can handle both regression and classification in a unified way. Another popular context-aware application is indoor localization \cite{li2019smartloc}. Multiple machine learning algorithms are trained, and their predictions are fused to improve accuracy. 

\vspace{0.05cm}\noindent{\bf{Caching:}}
Machine learning-based content caching methods have been proposed in \cite{lin2020cooperative,zhang2019double,sadeghi2017optimal}. One of the implementations is the intelligent base station with caching \cite{zhang2019double}. It has a placement delivery array system in the base station that uses the double-coded caching technique. It is formulated as an optimization problem that minimizes the delay and power consumption. The wireless network is modeled as an MDP with unknown transition probabilities because it is not available in a real-world scenario. Deep reinforcement learning is used to solve the MDP by taking the action of scheduling decisions and optimizing the reward of the transmission delay and power. 

\vspace{0.05cm}\noindent{\bf{Application Functionality and Management:}}
Traffic classification is another important application of AI techniques \cite{qazi2013application,testi2018machine}. The work in \cite{qazi2013application} developed Atlas on wireless networking at HP Labs. Atlas is a traffic classifier that can check the data traffic and identify its source software and applications. However, it is challenging to obtain training datasets for machine learning because massive and various network flow samples are hard to label. Atlas addresses this problem by using the mobile agents installed on some dedicated testing devices to collect the network logs, which are then used as the training data. 

Network function virtualization (NFV) is a key function in a software-defined network (SDN) \cite{shah2020multiagent,subramanya2019machine}. A software-defined radio was proposed in \cite{shah2020multiagent} to control IoT network parameters. NFV maps the transmission requests to virtual requests at the software level, which is modeled as MDP. MDP is solved by multi-agent deep reinforcement learning where every agent learns to select devices to form optimal routes and allocate proper power to devices. The study in \cite{luo2019adaptive} considered SDN management to support mobile edge clouds for video streaming. The functions of SDN such as video quality, transcoding, and caching are controlled by the virtual appliances of NFV. Both bandwidth allocation and power consumption of virtual appliances are optimized by deep reinforcement learning.

\vspace{0.05cm}\noindent{\bf{Computational Resource Management:}}
AI techniques can also help computational resource allocations in wireless networks \cite{zhang2019deep,tang2020decentralized,zhan2020deep,zeng2019resource}. In mobile edge computing, it is important to determine how to allocate the workload to mobile edges based on their available computing resources. A reinforcement learning-based framework for each edge to maximize each user's energy consumption and computing time is proposed in \cite{zhan2020deep}. The work of \cite{tang2020decentralized} considered an energy-saving model in IoT networks to reduce energy consumption based on a reinforcement learning model that allows every edge device to learn offloading decisions locally without accurate global information. 

\subsection{AI for Security}
We then review existing studies related to using AI techniques to enhance the wireless network performance. 

\subsubsection{Network Layer}
At the network layer, attacks mainly focus on disruptions to normal operations of network traffic. 

\vspace{0.05cm}\noindent{\bf{DoS Attacks:}}
The DoS attack is a common attack at the network layer \cite{kuadey2021deepsecure,almiani2021ddos, shamshirband2014cooperative}. 5G network slicing is a technology that divides a network into multiple virtual networks, which can be targeted by DoS attacks \cite{kuadey2021deepsecure}. A deep learning framework has been proposed in \cite{kuadey2021deepsecure} to jointly predict DoS attacks and slice traffic. The detection of DoS attacks is based on packet features including flow duration, internet protocol (IP) addresses, ports, and protocols. Deep learning with Kalman backpropagation was also proposed to detect DoS attacks in \cite{almiani2021ddos}. Features used in \cite{almiani2021ddos} include flow duration and flow inter-arrival time. The Kalman filter shows its capability to predict and detect DoS attacks by adjusting deep learning weights.


\vspace{0.05cm}\noindent{\bf{Loophole Attacks:}}
A new insider attack called loophole attack was proposed in \cite{chowdhury2021novel}. The attacker can be launched at a malicious gateway node. By intercepting and rerouting data in a loop to delay traffic, it can attack the IPv6 routing protocol for low-power and lossy networks. To counter attackers, traffic features such as rank, topology inconsistency, and rerouting procedures are used to train a deep learning framework. Simulations in \cite{chowdhury2021novel} show that the deep learning framework achieves more than $90\%$ accuracy to identify such attackers.

\vspace{0.05cm}\noindent{\bf{Anomaly Detection:}} 
Machine learning-based anomaly detection has become a common way to detect anomaly in network traffic based on packet features. A general comparison of machine learning-based anomaly detection was given in \cite{liu2020machine}, which tested various machine learning algorithms, including SVM, decision tree, random forest, and K-means; and common network attacks, including SYN flooding, land, UDP flood, ping of death, smurf, IP sweeping, and port scan. Generally, those machine learning algorithms can be used to detect suspicious features of network traffic. Tree-based methods were observed with better performance than others in \cite{liu2020machine}. 

\subsubsection{Application Layer}
Recently, there are also substantial efforts focusing on using AI for application layer security. 

\vspace{0.05cm}\noindent{\bf{Phishing and Malware:}}
A strategy against phishing in fog networks was designed in \cite{pham2018phishing} and built upon a neural network-based fuzzy detector. In the detector, 27 features are selected from uniform resource locator information and web information, and then are fuzzed as three classes and provided to a neural network to detect phishing. 
Machine learning has been proposed to detect malware at the application layer \cite{guizani2020network,amos2013applying,xiao2017cloud}. For example, Q-learning has been used in \cite{xiao2017cloud, guizani2020network} for malware detection in mobile and IoT networks, respectively. 

\vspace{0.05cm}\noindent{\bf{Location Privacy:}}
Location privacy \cite{fang2016virtual, wang2015location} has become an increasingly important topic recently with new attacks emerging to infer a mobile user's location. For example, \cite{han2022location} showed that attackers can target the application layer to steal the geographical information of users. The use of machine learning towards location privacy has been discussed in \cite{wang2020locjury,shaham2020privacy,papst2022share}. Anonymizing the spatiotemporal trajectory data is an effective method to protect privacy before publishing data \cite{shaham2020privacy}, in which trajectories are clustered by k-means to confuse adversaries without information loss measured by generalization hierarchy trees.


\vspace{0.05cm}\noindent{\bf{Cross-layer Defenses:}}
Some methods can work across layers to defend against attacks \cite{hussain2020deep,rawal2022identifying}. In \cite{hussain2020deep}, DDoS attacks across the PHY layer and application layers were taken into consideration. Three kinds of DDoS attacks were analyzed: silent call attacks, message spamming attacks, and signaling attacks, all leading to changes in network traffic features. A deep learning framework in \cite{hussain2020deep} was trained by a large volume of data to accurately detect such attacks. In \cite{rawal2022identifying}, distributed DoS of TCP, HTTP, and UDP protocols were considered. Decision trees were used to distinguish the features of flow because distributed DoS on different protocols will lead to some specific changes, such as the TCP SYN, HTTP GET, or POST requests.

\section{Summary of Challenges Going Forward}\label{sec:challenges}
Based on our review, we find that the application of AI on wireless networks is under rapid development, and there still exist challenges to be solved on the path ahead. 

\begin{itemize}
    \item Interpretability of operational wireless data. Many existing machine learning frameworks work like black boxes, which lack interpretability. They need experts to determine which features are dominant and should be used to train models \cite{chinchali2018cellular,goutay2020deep,lin2020contour}. How to select wireless network features, why these features are important, and accordingly lead to accurate classification for an AI-based design worth more research efforts. 
    \item Model Adaptation to Dynamics. A distinguishing feature of the wireless network is the dynamically-changing environmental data, such as user mobility and time-varying channel fading. They cause confusion, noise, and unreliability in data. Some models are designed typically to process certain kinds of data \cite{nachmani2016learning,teng2019low}. Generally, it is worthy of more studies regarding how a trained machine learning model based on wireless data for one environment for a time period can be reliably applied to a different environment at another time period. 
    \item Balancing between Complexity and Performance. Machine learning frameworks, in particular neural networks, can incur a higher complexity than conventional methods, indicating that IoT devices with limited cost budgets still have difficulty in adopting them. Low-complexity neural network implementation and deployment can provide one feasible way for AI-empowered IoT devices. 
    \item Balancing between AI and Conventional Methods. AI may not be the optimal choice for every wireless network task as conventional methods can provide more stable and interpretable results sometimes. Therefore, we think adequately adopting AI to balance between AI-based and conventional methods is important in wireless network operations. 
    \item Adversarial Machine Learning and Effective Defense. Using machine learning unfortunately creates a new dimension of security risks. Adversarial machine learning can attack existing machine learning models by maliciously manipulating the learning process with small perturbations \cite{luo2020attackers}. As a result, AI-based methods need to be carefully reviewed to address the risk of adversarial examples in wireless networking. 
\end{itemize}

\section{Conclusion}\label{sec:end}
In this paper, we surveyed the literature on a rapidly growing area of AI for wireless networking. We summarized the use of AI techniques from the PHY layer to the application layer in two major aspects: improving performance and enhancing security. We also discuss the challenges on the path ahead. As we have seen, different AI techniques can be applied or re-designed for various wireless algorithms, mechanisms, architectures, and systems, making AI for wireless networking a challenging and promising research area. 




\bibliographystyle{IEEEtran}
\bibliography{i3cn}

\begin{thebibliography}{100}
\providecommand{\url}[1]{#1}
\csname url@samestyle\endcsname
\providecommand{\newblock}{\relax}
\providecommand{\bibinfo}[2]{#2}
\providecommand{\BIBentrySTDinterwordspacing}{\spaceskip=0pt\relax}
\providecommand{\BIBentryALTinterwordstretchfactor}{4}
\providecommand{\BIBentryALTinterwordspacing}{\spaceskip=\fontdimen2\font plus
\BIBentryALTinterwordstretchfactor\fontdimen3\font minus
  \fontdimen4\font\relax}
\providecommand{\BIBforeignlanguage}[2]{{%
\expandafter\ifx\csname l@#1\endcsname\relax
\typeout{** WARNING: IEEEtran.bst: No hyphenation pattern has been}%
\typeout{** loaded for the language `#1'. Using the pattern for}%
\typeout{** the default language instead.}%
\else
\language=\csname l@#1\endcsname
\fi
#2}}
\providecommand{\BIBdecl}{\relax}
\BIBdecl

\bibitem{ye2019deep}
N.~Ye, X.~Li, H.~Yu, A.~Wang, W.~Liu, and X.~Hou, ``Deep learning aided
  grant-free {NOMA} toward reliable low-latency access in tactile {I}nternet of
  {T}hings,'' \emph{IEEE TII}, 2019.

\bibitem{santos2016intelligent}
J.~Santos, J.~J. Rodrigues, J.~Casal, K.~Saleem, and V.~Denisov, ``Intelligent
  personal assistants based on {Internet} of things approaches,'' \emph{IEEE
  Systems Journal}, 2016.

\bibitem{merchant2018deep}
K.~Merchant, S.~Revay, G.~Stantchev, and B.~Nousain, ``Deep learning for {RF}
  device fingerprinting in cognitive communication networks,'' \emph{IEEE
  JSTSP}, 2018.

\bibitem{wang2019data}
Y.~Wang, M.~Liu, J.~Yang, and G.~Gui, ``Data-driven deep learning for automatic
  modulation recognition in cognitive radios,'' \emph{IEEE TVT}, 2019.

\bibitem{xiong2021warmonger}
J.~Xiong, M.~Wei, Z.~Lu, and Y.~Liu, ``Warmonger: inflicting
  {D}enial-of-{S}ervice via serverless functions in the cloud,'' in \emph{ACM
  CCS}, 2021.

\bibitem{xu2017deep}
Z.~Xu, Y.~Wang, J.~Tang, J.~Wang, and M.~C. Gursoy, ``A deep reinforcement
  learning based framework for power-efficient resource allocation in cloud
  {RANs},'' in \emph{IEEE ICC}, 2017.

\bibitem{mehrabi2019decision}
M.~Mehrabi, M.~Mohammadkarimi, M.~Ardakani, and Y.~Jing, ``Decision directed
  channel estimation based on deep neural network k-step predictor for {MIMO}
  communications in {5G},'' \emph{IEEE JSAC}, 2019.

\bibitem{sadeghi2017optimal}
A.~Sadeghi, F.~Sheikholeslami, and G.~B. Giannakis, ``Optimal and scalable
  caching for {5G} using reinforcement learning of space-time popularities,''
  \emph{IEEE JSTSP}, 2017.

\bibitem{hampton2013introduction}
J.~R. Hampton, \emph{Introduction to {MIMO} communications}.\hskip 1em plus
  0.5em minus 0.4em\relax Cambridge university press, 2013.

\bibitem{yang2015fifty}
S.~Yang and L.~Hanzo, ``Fifty years of {MIMO} detection: The road to
  large-scale {MIMO}s,'' \emph{IEEE COMST}, 2015.

\bibitem{shlezinger2020deepsic}
N.~Shlezinger, R.~Fu, and Y.~C. Eldar, ``{DeepSIC}: Deep soft interference
  cancellation for multiuser {MIMO} detection,'' \emph{IEEE TWC}, 2020.

\bibitem{guo2020convolutional}
J.~Guo, C.-K. Wen, S.~Jin, and G.~Y. Li, ``Convolutional neural network-based
  multiple-rate compressive sensing for massive {MIMO} {CSI} feedback: Design,
  simulation, and analysis,'' \emph{IEEE TWC}, 2020.

\bibitem{teng2019low}
C.-F. Teng, C.-H.~D. Wu, A.~K.-S. Ho, and A.-Y.~A. Wu, ``Low-complexity
  recurrent neural network-based polar decoder with weight quantization
  mechanism,'' in \emph{IEEE ICASSP}, 2019.

\bibitem{xu2020application}
Y.~Xu, J.~Yu, and R.~M. Buehrer, ``The application of deep reinforcement
  learning to distributed spectrum access in dynamic heterogeneous environments
  with partial observations,'' \emph{IEEE TWC}, 2020.

\bibitem{xu2021deep}
J.~Xu and B.~Ai, ``Deep reinforcement learning for handover-aware {MPTCP}
  congestion control in space-ground integrated network of railways,''
  \emph{IEEE Wireless Commun.}, 2021.

\bibitem{wang2018efficient}
N.~Wang, L.~Jiao, P.~Wang, M.~Dabaghchian, and K.~Zeng, ``Efficient identity
  spoofing attack detection for {IoT} in mm-{W}ave and massive {MIMO} {5G}
  communication,'' in \emph{IEEE GLOBECOM}, 2018.

\bibitem{han2017two}
G.~Han, L.~Xiao, and H.~V. Poor, ``Two-dimensional anti-jamming communication
  based on deep reinforcement learning,'' in \emph{IEEE ICASSP}, 2017.

\bibitem{kuadey2021deepsecure}
N.~A.~E. Kuadey, G.~T. Maale, T.~Kwantwi, G.~Sun, and G.~Liu, ``Deepsecure:
  Detection of {D}istributed {D}enial of {S}ervice attacks on {5G} network
  slicing—deep learning approach,'' \emph{IEEE COMML}, 2021.

\bibitem{chinchali2018cellular}
S.~Chinchali, P.~Hu, T.~Chu, M.~Sharma, M.~Bansal, R.~Misra, M.~Pavone, and
  S.~Katti, ``Cellular network traffic scheduling with deep reinforcement
  learning,'' in \emph{AAAI Conference on AI}, 2018.

\bibitem{shah2020multiagent}
H.~A. Shah and L.~Zhao, ``Multiagent deep-reinforcement-learning-based virtual
  resource allocation through network function virtualization in {I}nternet of
  {T}hings,'' \emph{IEEE IoTM}, 2020.

\bibitem{mahesh2020machine}
B.~Mahesh, ``Machine learning algorithms-a review,'' \emph{IJSR}, 2020.

\bibitem{zhang2020taxonomy}
H.~Zhang and T.~Yu, \emph{Deep Reinforcement Learning: Fundamentals, Research
  and Applications}.\hskip 1em plus 0.5em minus 0.4em\relax Springer, 2020.

\bibitem{aggarwal2018neural}
C.~C. Aggarwal \emph{et~al.}, ``Neural networks and deep learning,''
  \emph{Springer}, 2018.

\bibitem{aoudia2019model}
F.~A. Aoudia and J.~Hoydis, ``Model-free training of end-to-end communication
  systems,'' \emph{IEEE JSAC}, 2019.

\bibitem{dorner2017deep}
S.~D{\"o}rner, S.~Cammerer, J.~Hoydis, and S.~Ten~Brink, ``Deep learning based
  communication over the air,'' \emph{IEEE JSTSP}, 2017.

\bibitem{raj2020design}
V.~Raj and S.~Kalyani, ``Design of communication systems using deep learning: A
  variational inference perspective,'' \emph{IEEE TCCN}, 2020.

\bibitem{goutay2020deep}
M.~Goutay, F.~A. Aoudia, and J.~Hoydis, ``Deep hypernetwork-based {MIMO}
  detection,'' in \emph{IEEE SPAWC}, 2020.

\bibitem{he2020model}
H.~He, C.-K. Wen, S.~Jin, and G.~Y. Li, ``Model-driven deep learning for {MIMO}
  detection,'' \emph{IEEE TSP}, 2020.

\bibitem{tan2018low}
X.~Tan, Z.~Zhong, Z.~Zhang, X.~You, and C.~Zhang, ``Low-complexity message
  passing {MIMO} detection algorithm with deep neural network,'' in \emph{IEEE
  GlobalSIP}, 2018.

\bibitem{chaudhari2020reliable}
S.~Chaudhari, H.~Kwon, and K.-B. Song, ``Reliable and low-complexity {MIMO}
  detector selection using neural network,'' in \emph{IEEE ICNC}, 2020.

\bibitem{balevi2020massive}
E.~Balevi, A.~Doshi, and J.~G. Andrews, ``Massive {MIMO} channel estimation
  with an untrained deep neural network,'' \emph{IEEE TWC}, 2020.

\bibitem{lu2020multi}
Z.~Lu, J.~Wang, and J.~Song, ``Multi-resolution {CSI} feedback with deep
  learning in massive {MIMO} system,'' in \emph{IEEE ICC}, 2020.

\bibitem{yang2020machine}
Y.~Yang, Z.~Gao, Y.~Ma, B.~Cao, and D.~He, ``Machine learning enabling analog
  beam selection for concurrent transmissions in millimeter-wave {V2V}
  communications,'' \emph{IEEE TVT}, 2020.

\bibitem{senigagliesi2020comparison}
L.~Senigagliesi, M.~Baldi, and E.~Gambi, ``Comparison of statistical and
  machine learning techniques for physical layer authentication,'' \emph{IEEE
  TIFS}, 2020.

\bibitem{fang2018learning}
H.~Fang, X.~Wang, and L.~Hanzo, ``Learning-aided physical layer authentication
  as an intelligent process,'' \emph{IEEE TCOM}, 2018.

\bibitem{xiao2016phy}
L.~Xiao, Y.~Li, G.~Han, G.~Liu, and W.~Zhuang, ``{PHY}-layer spoofing detection
  with reinforcement learning in wireless networks,'' \emph{IEEE TVT}, 2016.

\bibitem{xiao2017phy}
L.~Xiao, X.~Wan, and Z.~Han, ``{PHY}-layer authentication with multiple
  landmarks with reduced overhead,'' \emph{IEEE TWC}, 2017.

\bibitem{senigagliesi2019statistical}
L.~Senigagliesi, M.~Baldi, and E.~Gambi, ``Statistical and machine
  learning-based decision techniques for physical layer authentication,'' in
  \emph{IEEE GLOBECOM}, 2019.

\bibitem{jiang2018virtual}
P.~Jiang, H.~Wu, C.~Wang, and C.~Xin, ``Virtual {MAC} spoofing detection
  through deep learning,'' in \emph{IEEE ICC}, 2018.

\bibitem{benzaid2016intelligent}
C.~Benza{\"\i}d, A.~Boulgheraif, F.~Z. Dahmane, A.~Al-Nemrat, and K.~Zeraoulia,
  ``Intelligent detection of {MAC} spoofing attack in 802.11 network,'' in
  \emph{ACM ICDCN}, 2016.

\bibitem{zhao2020comb}
S.~Zhao, Z.~Qu, Z.~Luo, Z.~Lu, and Y.~Liu, ``Comb decoding towards
  collision-free {WiFi},'' in \emph{USENIX NSDI}, 2020.

\bibitem{nachmani2016learning}
E.~Nachmani, Y.~Be'ery, and D.~Burshtein, ``Learning to decode linear codes
  using deep learning,'' in \emph{IEEE Allerton}, 2016.

\bibitem{huang2021deep}
Y.~Huang, Y.~T. Hou, and W.~Lou, ``A deep-learning-based link adaptation design
  for e{MBB}/{URLLC} multiplexing in {5G} {NR},'' in \emph{IEEE INFOCOM}, 2021.

\bibitem{saxena2021reinforcement}
V.~Saxena, H.~Tullberg, and J.~Jald{\'e}n, ``Reinforcement learning for
  efficient and tuning-free link adaptation,'' \emph{IEEE TWC}, 2021.

\bibitem{he2018model}
H.~He, C.-K. Wen, S.~Jin, and G.~Y. Li, ``A model-driven deep learning network
  for {MIMO} detection,'' in \emph{IEEE GlobalSIP}, 2018.

\bibitem{khani2020adaptive}
M.~Khani, M.~Alizadeh, J.~Hoydis, and P.~Fleming, ``Adaptive neural signal
  detection for massive {MIMO},'' \emph{IEEE TWC}, 2020.

\bibitem{he2018deep}
H.~He, C.-K. Wen, S.~Jin, and G.~Y. Li, ``Deep learning-based channel
  estimation for beamspace {mmWave} massive {MIMO} systems,'' \emph{IEEE
  COMML}, 2018.

\bibitem{liu2020deep}
S.~Liu, Z.~Gao, J.~Zhang, M.~Di~Renzo, and M.-S. Alouini, ``Deep denoising
  neural network assisted compressive channel estimation for {mmWave}
  intelligent reflecting surfaces,'' \emph{IEEE TVT}, 2020.

\bibitem{wang2018deep}
T.~Wang, C.-K. Wen, S.~Jin, and G.~Y. Li, ``Deep learning-based {CSI} feedback
  approach for time-varying massive {MIMO} channels,'' \emph{IEEE COMML}, 2018.

\bibitem{wen2018deep}
C.-K. Wen, W.-T. Shih, and S.~Jin, ``Deep learning for massive {MIMO} {CSI}
  feedback,'' \emph{IEEE WCL}, 2018.

\bibitem{lin2020contour}
Y.~Lin, Y.~Tu, Z.~Dou, L.~Chen, and S.~Mao, ``Contour stella image and deep
  learning for signal recognition in the physical layer,'' \emph{IEEE TCCN},
  2020.

\bibitem{aslam2012automatic}
M.~W. Aslam, Z.~Zhu, and A.~K. Nandi, ``Automatic modulation classification
  using combination of genetic programming and knn,'' \emph{IEEE TWC}, 2012.

\bibitem{ahn2022machine}
H.~Ahn, I.~Orikumhi, J.~Kang, H.~Park, H.~Jwa, J.~Na, and S.~Kim, ``Machine
  learning-based vision-aided beam selection for {mmWave} multiuser {MISO}
  system,'' \emph{IEEE COMML}, 2022.

\bibitem{klautau20185g}
A.~Klautau, P.~Batista, N.~Gonz{\'a}lez-Prelcic, Y.~Wang, and R.~W. Heath,
  ``{5G} {MIMO} data for machine learning: Application to beam-selection using
  deep learning,'' in \emph{IEEE ITA}, 2018.

\bibitem{chiang2021machine}
H.-L. Chiang, K.-C. Chen, W.~Rave, M.~K. Marandi, and G.~Fettweis,
  ``Machine-learning beam tracking and weight optimization for {mmWave}
  multi-{UAV} links,'' \emph{IEEE TWC}, 2021.

\bibitem{lim2021deep}
S.~H. Lim, S.~Kim, B.~Shim, and J.~W. Choi, ``Deep learning-based beam tracking
  for millimeter-wave communications under mobility,'' \emph{IEEE TCOM}, 2021.

\bibitem{sarkar2021machine}
S.~Sarkar, M.~Krunz, I.~Aykin, and D.~Manzi, ``Machine learning for robust beam
  tracking in mobile millimeter-wave systems,'' in \emph{IEEE GLOBECOM}, 2021.

\bibitem{heng2021machine}
Y.~Heng and J.~G. Andrews, ``Machine learning-assisted beam alignment for
  {mmWave} systems,'' \emph{IEEE TCCN}, 2021.

\bibitem{ma2020machine}
W.~Ma, C.~Qi, and G.~Y. Li, ``Machine learning for beam alignment in millimeter
  wave massive {MIMO},'' \emph{IEEE COMML}, 2020.

\bibitem{domae2021machine}
B.~W. Domae, R.~Li, and D.~Cabric, ``Machine learning assisted phase-less
  millimeter-wave beam alignment in multipath channels,'' in \emph{IEEE
  GLOBECOM}, 2021.

\bibitem{meng2018deep}
X.~Meng, H.~Inaltekin, and B.~Krongold, ``Deep reinforcement learning-based
  power control in full-duplex cognitive radio networks,'' in \emph{IEEE
  GLOBECOM}, 2018.

\bibitem{mismar2019framework}
F.~B. Mismar, J.~Choi, and B.~L. Evans, ``A framework for automated cellular
  network tuning with reinforcement learning,'' \emph{IEEE TCOM}, 2019.

\bibitem{lee2019resource}
H.-S. Lee, J.-Y. Kim, and J.-W. Lee, ``Resource allocation in wireless networks
  with deep reinforcement learning: A circumstance-independent approach,''
  \emph{IEEE Systems Journal}, 2019.

\bibitem{zhang2020power}
H.~Zhang, N.~Yang, W.~Huangfu, K.~Long, and V.~C. Leung, ``Power control based
  on deep reinforcement learning for spectrum sharing,'' \emph{IEEE TWC}, 2020.

\bibitem{nasir2019multi}
Y.~S. Nasir and D.~Guo, ``Multi-agent deep reinforcement learning for dynamic
  power allocation in wireless networks,'' \emph{IEEE JSAC}, 2019.

\bibitem{xiao2019reinforcement}
L.~Xiao, H.~Zhang, Y.~Xiao, X.~Wan, S.~Liu, L.-C. Wang, and H.~V. Poor,
  ``Reinforcement learning-based downlink interference control for ultra-dense
  small cells,'' \emph{IEEE TWC}, 2019.

\bibitem{liu2018deepnap}
J.~Liu, B.~Krishnamachari, S.~Zhou, and Z.~Niu, ``Deep{N}ap: Data-driven base
  station sleeping operations through deep reinforcement learning,'' \emph{IEEE
  IoTM}, 2018.

\bibitem{wang2018deepmultichannel}
S.~Wang, H.~Liu, P.~H. Gomes, and B.~Krishnamachari, ``Deep reinforcement
  learning for dynamic multichannel access in wireless networks,'' \emph{IEEE
  TCCN}, 2018.

\bibitem{naparstek2018deep}
O.~Naparstek and K.~Cohen, ``Deep multi-user reinforcement learning for
  distributed dynamic spectrum access,'' \emph{IEEE TWC}, 2018.

\bibitem{challita2018proactive}
U.~Challita, L.~Dong, and W.~Saad, ``Proactive resource management for {LTE} in
  unlicensed spectrum: A deep learning perspective,'' \emph{IEEE TWC}, 2018.

\bibitem{li2019multi}
Z.~Li and C.~Guo, ``Multi-agent deep reinforcement learning based spectrum
  allocation for {D2D} underlay communications,'' \emph{IEEE TVT}, 2019.

\bibitem{koushik2017intelligent}
A.~Koushik, F.~Hu, and S.~Kumar, ``Intelligent spectrum management based on
  transfer actor-critic learning for rateless transmissions in cognitive radio
  networks,'' \emph{IEEE TMC}, 2017.

\bibitem{li2020deep}
Y.~Li, W.~Zhang, C.-X. Wang, J.~Sun, and Y.~Liu, ``Deep reinforcement learning
  for dynamic spectrum sensing and aggregation in multi-channel wireless
  networks,'' \emph{IEEE TCCN}, 2020.

\bibitem{ye2020deepnoma}
N.~Ye, X.~Li, H.~Yu, L.~Zhao, W.~Liu, and X.~Hou, ``{DeepNOMA}: A unified
  framework for {NOMA} using deep multi-task learning,'' \emph{IEEE TWC}, 2020.

\bibitem{zhong2021ai}
R.~Zhong, Y.~Liu, X.~Mu, Y.~Chen, and L.~Song, ``{AI} empowered {RIS}-assisted
  {NOMA} networks: Deep learning or reinforcement learning?'' \emph{IEEE JSAC},
  2021.

\bibitem{wang2021joint}
S.~Wang, T.~Lv, W.~Ni, N.~C. Beaulieu, and Y.~J. Guo, ``Joint resource
  management for {MC-NOMA}: A deep reinforcement learning approach,''
  \emph{IEEE TWC}, 2021.

\bibitem{rupasinghe2015reinforcement}
N.~Rupasinghe and {\.I}.~G{\"u}ven{\c{c}}, ``Reinforcement learning for
  licensed-assisted access of {LTE} in the unlicensed spectrum,'' in \emph{IEEE
  WCNC}, 2015.

\bibitem{nisioti2019robust}
E.~Nisioti and N.~Thomos, ``Robust coordinated reinforcement learning for {MAC}
  design in sensor networks,'' \emph{IEEE JSAC}, 2019.

\bibitem{nisioti2019fast}
------, ``Fast {Q}-learning for improved finite length performance of irregular
  repetition slotted {ALOHA},'' \emph{IEEE TCCN}, 2019.

\bibitem{sana2019multi}
M.~Sana, A.~De~Domenico, and E.~C. Strinati, ``Multi-agent deep reinforcement
  learning based user association for dense {mmWave} networks,'' in \emph{IEEE
  GLOBECOM}, 2019.

\bibitem{zhao2018deep}
N.~Zhao, Y.-C. Liang, D.~Niyato, Y.~Pei, and Y.~Jiang, ``Deep reinforcement
  learning for user association and resource allocation in heterogeneous
  networks,'' in \emph{IEEE GLOBECOM}, 2018.

\bibitem{zhao2019deep}
N.~Zhao, Y.-C. Liang, D.~Niyato, Y.~Pei, M.~Wu, and Y.~Jiang, ``Deep
  reinforcement learning for user association and resource allocation in
  heterogeneous cellular networks,'' \emph{IEEE TWC}, 2019.

\bibitem{hu2021smartphone}
C.~Hu, Y.~Liu, Z.~Lu, S.~Zhao, X.~Han, and J.~Xiong, ``Smartphone location
  spoofing attack in wireless networks,'' in \emph{SecureComm}, 2021.

\bibitem{shi2020generative}
Y.~Shi, K.~Davaslioglu, and Y.~E. Sagduyu, ``Generative adversarial network in
  the air: Deep adversarial learning for wireless signal spoofing,'' \emph{IEEE
  TCCN}, 2020.

\bibitem{xiao2015spoofing}
L.~Xiao, Y.~Li, G.~Liu, Q.~Li, and W.~Zhuang, ``Spoofing detection with
  reinforcement learning in wireless networks,'' in \emph{IEEE GLOBECOM}, 2015.

\bibitem{alagil2016randomized}
A.~Alagil, M.~Alotaibi, and Y.~Liu, ``Randomized positioning {DSSS} for
  anti-jamming wireless communications,'' in \emph{IEEE ICNC}, 2016.

\bibitem{wu2011anti}
Y.~Wu, B.~Wang, K.~R. Liu, and T.~C. Clancy, ``Anti-jamming games in
  multi-channel cognitive radio networks,'' \emph{IEEE JSAC}, 2011.

\bibitem{labib2015colonel}
M.~Labib, S.~Ha, W.~Saad, and J.~H. Reed, ``A colonel blotto game for
  anti-jamming in the {Internet} of things,'' in \emph{IEEE GLOBECOM}, 2015.

\bibitem{xiao2015user}
L.~Xiao, J.~Liu, Q.~Li, N.~B. Mandayam, and H.~V. Poor, ``User-centric view of
  jamming games in cognitive radio networks,'' \emph{IEEE TIFS}, 2015.

\bibitem{erpek2018deep}
T.~Erpek, Y.~E. Sagduyu, and Y.~Shi, ``Deep learning for launching and
  mitigating wireless jamming attacks,'' \emph{IEEE TCCN}, 2018.

\bibitem{gwon2013competing}
Y.~Gwon, S.~Dastangoo, C.~Fossa, and H.~Kung, ``Competing mobile network game:
  Embracing antijamming and jamming strategies with reinforcement learning,''
  in \emph{IEEE CNS}, 2013.

\bibitem{aref2017multi}
M.~A. Aref, S.~K. Jayaweera, and S.~Machuzak, ``Multi-agent reinforcement
  learning based cognitive anti-jamming,'' in \emph{IEEE WCNC}, 2017.

\bibitem{letafati2021deep}
M.~Letafati, H.~Behroozi, B.~H. Khalaj, and E.~A. Jorswieck, ``Deep learning
  for hardware-impaired wireless secret key generation with man-in-the-middle
  attacks,'' in \emph{IEEE GLOBECOM}, 2021.

\bibitem{fang2016mimicry}
S.~Fang, Y.~Liu, and P.~Ning, ``Mimicry attacks against wireless link signature
  and new defense using time-synched link signature,'' \emph{IEEE TIFS},
  vol.~11, 2016.

\bibitem{agarwal2015detection}
M.~Agarwal, S.~Biswas, and S.~Nandi, ``Detection of de-authentication {DoS}
  attacks in {Wi-Fi} networks: A machine learning approach,'' in \emph{IEEE
  SMC}, 2015.

\bibitem{luo2020adversarial}
Z.~Luo, S.~Zhao, Z.~Lu, Y.~E. Sagduyu, and J.~Xu, ``Adversarial machine
  learning based partial-model attack in {IoT},'' in \emph{ACM WiseML}, 2020.

\bibitem{luo2020attackers}
Z.~Luo, S.~Zhao, Z.~Lu, J.~Xu, and Y.~E. Sagduyu, ``When attackers meet {AI}:
  Learning-empowered attacks in cooperative spectrum sensing,'' \emph{IEEE
  TMC}, 2020.

\bibitem{luo2021low}
Z.~Luo, S.~Zhao, R.~Duan, Z.~Lu, Y.~E. Sagduyu, and J.~Xu, ``Low-cost
  influence-limiting defense against adversarial machine learning attacks in
  cooperative spectrum sensing,'' in \emph{ACM WiseML}, 2021.

\bibitem{kavousi2020machine}
A.~Kavousi-Fard, W.~Su, and T.~Jin, ``A machine-learning-based cyber attack
  detection model for wireless sensor networks in microgrids,'' \emph{IEEE
  TII}, 2020.

\bibitem{han2021anomaly}
G.~Han, J.~Tu, L.~Liu, M.~Martinez-Garcia, and Y.~Peng, ``Anomaly detection
  based on multidimensional data processing for protecting vital devices in
  {6G}-enabled massive {IIoT},'' \emph{IEEE IoTM}, 2021.

\bibitem{satam2020wids}
P.~Satam and S.~Hariri, ``{WIDS}: An anomaly based intrusion detection system
  for {Wi-Fi} ({IEEE} 802.11) protocol,'' \emph{IEEE TNSM}, 2020.

\bibitem{meng2019revealing}
Y.~Meng, J.~Li, H.~Zhu, X.~Liang, Y.~Liu, and N.~Ruan, ``Revealing your mobile
  password via {WiFi} signals: Attacks and countermeasures,'' \emph{IEEE TMC},
  2019.

\bibitem{fang2018no}
S.~Fang, I.~Markwood, Y.~Liu, S.~Zhao, Z.~Lu, and H.~Zhu, ``No training
  hurdles: Fast training-agnostic attacks to infer your typing,'' in \emph{ACM
  CCS}, 2018.

\bibitem{he2018learning}
D.~He, C.~Liu, H.~Wang, and T.~Q. Quek, ``Learning-based wireless powered
  secure transmission,'' \emph{IEEE COMML}, 2018.

\bibitem{vashist2019securing}
A.~Vashist, A.~Keats, S.~M.~P. Dinakarrao, and A.~Ganguly, ``Securing a
  wireless network-on-chip against jamming-based {D}enial-of-{S}ervice and
  eavesdropping attacks,'' \emph{IEEE TVLSI Systems}, 2019.

\bibitem{testi2018machine}
E.~Testi, E.~Favarelli, and A.~Giorgetti, ``Machine learning for user traffic
  classification in wireless systems,'' in \emph{IEEE EUSIPCO}, 2018.

\bibitem{ye2009optimal}
Z.~Ye, A.~A. Abouzeid, and J.~Ai, ``Optimal stochastic policies for distributed
  data aggregation in wireless sensor networks,'' \emph{IEEE/ACM TON}, 2009.

\bibitem{yu2011adaptive}
B.~Yu, C.-Z. Xu, and M.~Guo, ``Adaptive forwarding delay control for {VANET}
  data aggregation,'' \emph{IEEE TPDS}, 2011.

\bibitem{taherkhani2016centralized}
N.~Taherkhani and S.~Pierre, ``Centralized and localized data congestion
  control strategy for vehicular ad hoc networks using a machine learning
  clustering algorithm,'' \emph{IEEE T-ITS}, 2016.

\bibitem{gholipour2017hop}
M.~Gholipour, A.~T. Haghighat, and M.~R. Meybodi, ``Hop-by-hop congestion
  avoidance in wireless sensor networks based on genetic support vector
  machine,'' \emph{Neurocomputing}, 2017.

\bibitem{al2011end}
A.~A. Al~Islam and V.~Raghunathan, ``End-to-end congestion control in wireless
  mesh networks using a neural network,'' in \emph{2011 IEEE WCNC}, 2011.

\bibitem{li2018qtcp}
W.~Li, F.~Zhou, K.~R. Chowdhury, and W.~Meleis, ``{QTCP}: Adaptive congestion
  control with reinforcement learning,'' \emph{IEEE TNSE}, 2018.

\bibitem{donta2020congestion}
P.~K. Donta, T.~Amgoth, and C.~S.~R. Annavarapu, ``Congestion-aware data
  acquisition with {Q}-learning for wireless sensor networks,'' in \emph{IEEE
  IEMTRONICS}, 2020.

\bibitem{cui2020improving}
L.~Cui, Z.~Yuan, Z.~Ming, and S.~Yang, ``Improving the congestion control
  performance for mobile networks in high-speed railway via deep reinforcement
  learning,'' \emph{IEEE TVT}, 2020.

\bibitem{he2021deepcc}
B.~He, J.~Wang, Q.~Qi, H.~Sun, J.~Liao, C.~Du, X.~Yang, and Z.~Han, ``{DeepCC}:
  Multi-agent deep reinforcement learning congestion control for multi-path
  {TCP} based on self-attention,'' \emph{IEEE TNSM}, 2021.

\bibitem{xie2019adaptive}
R.~Xie, X.~Jia, and K.~Wu, ``Adaptive online decision method for initial
  congestion window in {5G} mobile edge computing using deep reinforcement
  learning,'' \emph{IEEE JSAC}, 2019.

\bibitem{yao2017deepsense}
S.~Yao, S.~Hu, Y.~Zhao, A.~Zhang, and T.~Abdelzaher, ``Deepsense: A unified
  deep learning framework for time-series mobile sensing data processing,'' in
  \emph{ACM WWW}, 2017.

\bibitem{li2019smartloc}
L.~Li, X.~Guo, and N.~Ansari, ``{SmartLoc}: Smart wireless indoor localization
  empowered by machine learning,'' \emph{IEEE TIE}, 2019.

\bibitem{wang2016device}
J.~Wang, X.~Zhang, Q.~Gao, H.~Yue, and H.~Wang, ``Device-free wireless
  localization and activity recognition: A deep learning approach,'' \emph{IEEE
  TVT}, 2016.

\bibitem{vashist2020indoor}
A.~Vashist, D.~R. Bhanushali, R.~Relyea, C.~Hochgraf, A.~Ganguly, P.~S. Manoj,
  R.~Ptucha, A.~Kwasinski, and M.~E. Kuhl, ``Indoor wireless localization using
  consumer-grade 60 {GHz} equipment with machine learning for intelligent
  material handling,'' in \emph{IEEE ICCE}, 2020.

\bibitem{lin2020cooperative}
P.~Lin, Q.~Song, J.~Song, A.~Jamalipour, and F.~R. Yu, ``Cooperative caching
  and transmission in {CoMP}-integrated cellular networks using reinforcement
  learning,'' \emph{IEEE TVT}, 2020.

\bibitem{zhang2019double}
Z.~Zhang, H.~Chen, M.~Hua, C.~Li, Y.~Huang, and L.~Yang, ``Double coded caching
  in ultra dense networks: Caching and multicast scheduling via deep
  reinforcement learning,'' \emph{IEEE TCOM}, 2019.

\bibitem{qazi2013application}
Z.~A. Qazi, J.~Lee, T.~Jin, G.~Bellala, M.~Arndt, and G.~Noubir,
  ``Application-awareness in {SDN},'' in \emph{ACM SIGCOMM}, 2013.

\bibitem{subramanya2019machine}
T.~Subramanya and R.~Riggio, ``Machine learning-driven scaling and placement of
  virtual network functions at the network edges,'' in \emph{IEEE NetSoft},
  2019.

\bibitem{luo2019adaptive}
J.~Luo, F.~R. Yu, Q.~Chen, and L.~Tang, ``Adaptive video streaming with edge
  caching and video transcoding over software-defined mobile networks: A deep
  reinforcement learning approach,'' \emph{IEEE TWC}, 2019.

\bibitem{zhang2019deep}
H.~Zhang, W.~Wu, C.~Wang, M.~Li, and R.~Yang, ``Deep reinforcement
  learning-based offloading decision optimization in mobile edge computing,''
  in \emph{IEEE WCNC}, 2019.

\bibitem{tang2020decentralized}
Q.~Tang, R.~Xie, F.~R. Yu, T.~Huang, and Y.~Liu, ``Decentralized computation
  offloading in {IoT} fog computing system with energy harvesting: A
  {Dec-POMDP} approach,'' \emph{IEEE IoTM}, 2020.

\bibitem{zhan2020deep}
Y.~Zhan, S.~Guo, P.~Li, and J.~Zhang, ``A deep reinforcement learning based
  offloading game in edge computing,'' \emph{IEEE TC}, 2020.

\bibitem{zeng2019resource}
D.~Zeng, L.~Gu, S.~Pan, J.~Cai, and S.~Guo, ``Resource management at the
  network edge: A deep reinforcement learning approach,'' \emph{IEEE Network},
  2019.

\bibitem{almiani2021ddos}
M.~Almiani, A.~AbuGhazleh, Y.~Jararweh, and A.~Razaque, ``{DDoS} detection in
  {5G}-enabled {IoT} networks using deep {Kalman} backpropagation neural
  network,'' \emph{Springer IJMLC}, 2021.

\bibitem{shamshirband2014cooperative}
S.~Shamshirband, A.~Patel, N.~B. Anuar, M.~L.~M. Kiah, and A.~Abraham,
  ``Cooperative game theoretic approach using fuzzy {Q}-learning for detecting
  and preventing intrusions in wireless sensor networks,'' \emph{Elsevier
  Engineering Applications of Artificial Intelligence}, 2014.

\bibitem{chowdhury2021novel}
M.~Chowdhury, B.~Ray, S.~Chowdhury, and S.~Rajasegarar, ``A novel insider
  attack and machine learning based detection for the {Internet of Things},''
  \emph{ACM TIOT}, 2021.

\bibitem{liu2020machine}
J.~Liu, B.~Kantarci, and C.~Adams, ``Machine learning-driven intrusion
  detection for {Contiki-NG-based} {IoT} networks exposed to {NSL-KDD}
  dataset,'' in \emph{ACM WiseML}, 2020.

\bibitem{pham2018phishing}
C.~Pham, L.~A. Nguyen, N.~H. Tran, E.-N. Huh, and C.~S. Hong, ``Phishing-aware:
  A neuro-fuzzy approach for anti-phishing on fog networks,'' \emph{IEEE TNSM},
  2018.

\bibitem{guizani2020network}
N.~Guizani and A.~Ghafoor, ``A network function virtualization system for
  detecting malware in large {IoT} based networks,'' \emph{IEEE JSAC}, 2020.

\bibitem{amos2013applying}
B.~Amos, H.~Turner, and J.~White, ``Applying machine learning classifiers to
  dynamic android malware detection at scale,'' in \emph{IEEE IWCMC}, 2013.

\bibitem{xiao2017cloud}
L.~Xiao, Y.~Li, X.~Huang, and X.~Du, ``Cloud-based malware detection game for
  mobile devices with offloading,'' \emph{IEEE TMC}, 2017.

\bibitem{fang2016virtual}
S.~Fang, Y.~Liu, W.~Shen, H.~Zhu, and T.~Wang, ``Virtual multipath attack and
  defense for location distinction in wireless networks,'' \emph{IEEE TMC},
  2016.

\bibitem{wang2015location}
T.~Wang, Y.~Liu, Q.~Pei, and T.~Hou, ``Location-restricted services access
  control leveraging pinpoint waveforming,'' in \emph{ACM CCS}, 2015.

\bibitem{han2022location}
X.~Han, J.~Xiong, W.~Shen, Z.~Lu, and Y.~Liu, ``Location heartbleeding: The
  rise of {Wi-Fi} spoofing attack via geolocation {API},'' in \emph{ACM CCS},
  2022.

\bibitem{wang2020locjury}
Y.~Wang, Z.~Tian, Y.~Sun, X.~Du, and N.~Guizani, ``{LocJury}: an {IBN}-based
  location privacy preserving scheme for {IoCV},'' \emph{IEEE TITS}, 2020.

\bibitem{shaham2020privacy}
S.~Shaham, M.~Ding, B.~Liu, S.~Dang, Z.~Lin, and J.~Li, ``Privacy preserving
  location data publishing: A machine learning approach,'' \emph{IEEE T-ITS},
  2020.

\bibitem{papst2022share}
F.~Papst, N.~Stricker, R.~Entezari, and O.~Saukh, ``To share or not to share:
  On location privacy in {IoT} sensor data,'' in \emph{IEEE/ACM IoTDI}, 2022.

\bibitem{hussain2020deep}
B.~Hussain, Q.~Du, B.~Sun, and Z.~Han, ``Deep learning-based {DDoS}-attack
  detection for cyber--physical system over {5G} network,'' \emph{IEEE TII},
  2020.

\bibitem{rawal2022identifying}
B.~S. Rawal, S.~Patel, and M.~Sathiyanarayanan, ``Identifying {DDoS} attack
  using split-machine learning system in {5G} and beyond networks,'' in
  \emph{IEEE INFOCOM WKSHPS}, 2022.

\end{thebibliography}

\end{document}